\documentclass[12pt]{article}
\usepackage[paper=letterpaper,top=1in,bottom=1in,left=1in,right=1in]{geometry}
\usepackage{color,pgf}
\usepackage{epsfig,subfigure}

\newcommand{\Otrack}{Opti-Track}
\newcommand{\Keywords}[1]{\vskip1em plus6pt minus4pt \hspace{1em}{\bfseries Keywords: } #1\hfill}
\title{Sensor Noise Rejection}

\author{Rakshit Allamraju \and Ben Reish\and Allan Axelrod}
\date{}

\begin{document}
\maketitle
\begin{abstract}
    An inaccessible control architecture caused an undesirable influence on a UAV.  The encountered noise in the performance was modeled using stochastic methods and a corrective term was implemented on an external controller. Our findings suggest that the sonar noise problem is unconventional may warrant the development of a new methodology. 
    
    \Keywords{Subsumption Architecture,Gaussian Mixture Models,Gaussian Process, Quadcopter, Random Variable}
\end{abstract}

\section{Introduction}
Industrial applications of controlled solutions vary widely in the level of
 documentation made available to the end-users.  In some instances, entire
 portions of the software for a controlled solution are designed to be 
 inaccessible.  This limits the
 usefulness of purchased controlled solutions to certain niches that may not
 adequately encompass the scope of an engineering project \cite{Murphy2010}.  In such a case,
 projects may be delayed until an entirely new controlled solution is purchased,
 shipped, activated and reprogrammed.  However, project delays could result in
 unacceptable costs; e.g., in emergency response scenarios,  and delays in implementation could result in the 
 unnecessary loss of human lives \cite{Murphy2008,Murphy2009}. 
 
 Additionally, the increasingly aggressive nature of cyber-attacks has posed a
 serious security and safety concern .  In fact, the power, 
 automation and chemical industries have already been the targets of 
 debilitating cyber-attacks \cite{Kundur2010,Lewis2002}.  The cyber-warfare arms race
 has evolved in a costly and overly complicated manner,
 often involving efforts related to securing a network of systems with many
 accessible input sites.  Such efforts typically involve a trade off between
 operational efficiency and security. 
   
 In response to the aforementioned issues, we present an augmented control
 architecture that may similarly augment commercial-off-the-shelf 
 (COTS) hardware and potential cyber-warfare targets.  The 
 augmented architecture enlists an external sensory system to
 characterize temporally anomalous or otherwise undesirable behavior to shape
 the system input signal such that the undesirable aspects of the system 
 behavior are canceled out.  In the context of COTS systems, this may allow for
 systems to operate in more exotic environments and fulfill a broader range of
 operational capabilities.  For cyber-warefare targets, this solution may
 allow systems to continue to exhibit desirable performance before, during and 
 after a cyber-attack

 For the sake of simplicity, we experimentally demonstrate the effectiveness of
 the augmented control architecture in the context of COTS systems.  The 
 demonstrated noise rejection is notably different than what we've seen in literature and this report is an initial survey of our investigation.
 
\section{Problem Formulation}

\pgfdeclareimage[width = 0.46\textwidth]{Zsserrorflat}{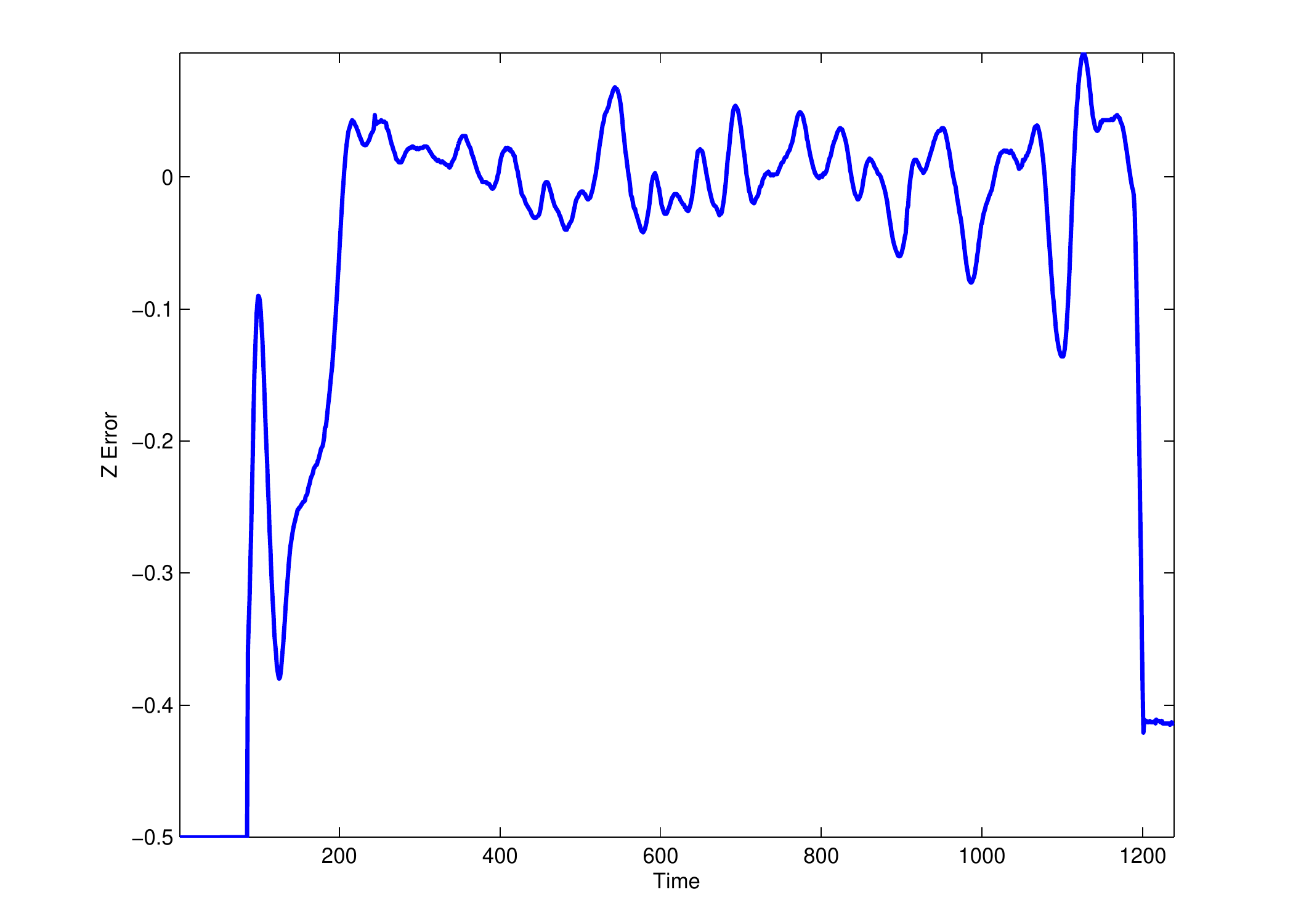}
\pgfdeclareimage[width = 0.46\textwidth]{Zsserroredge}{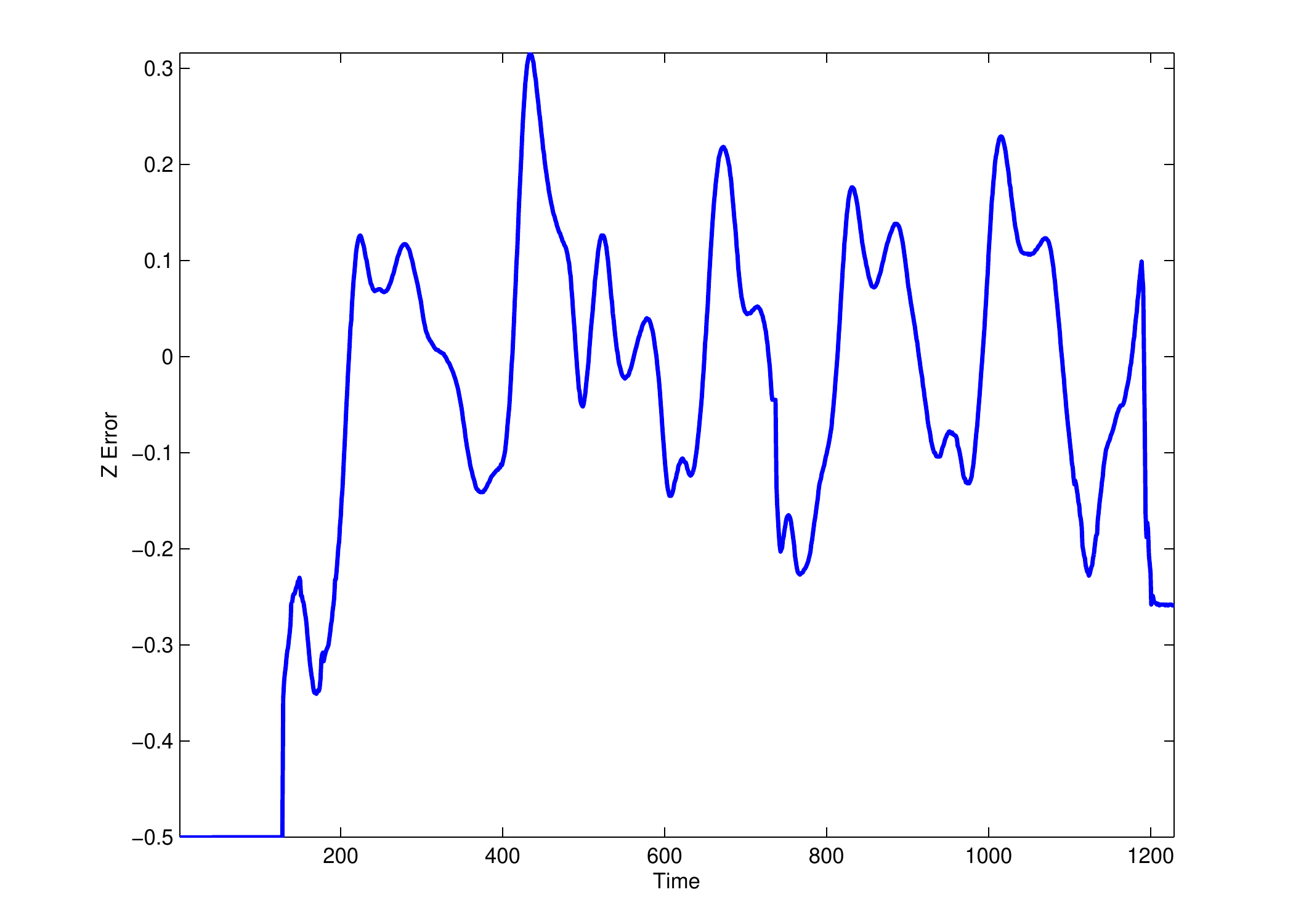}

\begin{figure}
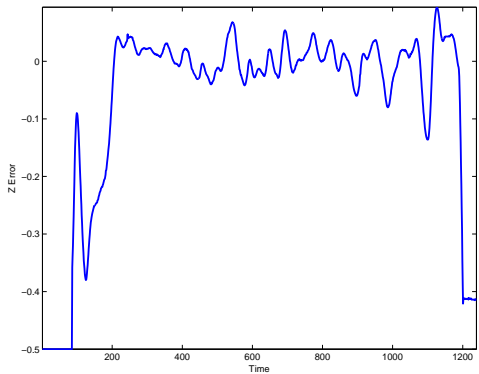
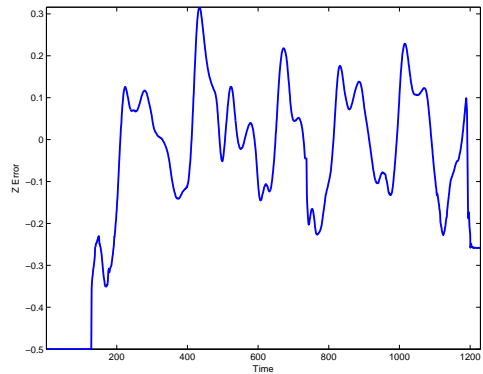

\subfigure[ZError -No Building]{\pgfuseimage{Zsserrorflat}

}
\subfigure[ZError -Building]{\pgfuseimage{Zsserroredge}
}
\caption{Zerror Comparison}

\label{fig:Steady State Error}
\end{figure}

The objective is to characterize and reduce the undesirable input (noise) introduced into the control input by a lower-level control loop of the subsumption architecture \cite{Brooks} as the UAV hovers about a location in 3-dimensional space.  The UAV has an internal
controller that uses a sonar apparatus to maintain a preset height of $1 m$.  
An \Otrack{} system is used as an external sensor to measure the UAV position during each flight testing episode.  The data from the \Otrack{} system is received by a Robot Operating System (ROS) client, which contains the higher-level control loop of the subsumption architecture.  The ROS client is used to affect the position and orientation of the UAV.  Since the default UAV altitude is above the observable range of the \Otrack, the ROS client altitude controller enforces an alternative steady state altitude of $0.5 m $.  The composite effect on the UAV altitude as it hovers over a uniformly level surface is shown in Figure \ref{fig:Steady State Error}a.  The composite effect on the UAV altitude as it hovers over a stepped surface is shown in Figure \ref{fig:Steady State Error}b.
  The degradation of the composite UAV altitude control in the presence of the stepped surface is significant and undesirable.

The approach we have considered for fitting our data are described below.

\subsection{Gaussian Mixture Models} 
Gaussian mixture models (GMM's) are one of the most mature approaches in modeling density estimation 
of data. GMM's are parametric probability distributions which are represented as weighted sums of 
Gaussian component densities. GMM's are employed for multiple purposes such as color-based tracking 
and segmentation, classifying color textures in image features \cite{Permuter}, and speaker 
recognition \cite{Reynolds}. A GMM can be represented as 
\begin{equation}
p(x|\lambda) = \sum\limits_{i = 1}^{M} w_i g_i(x_i|\mu_i,\sigma_i)
\end{equation}
GMM's add a great deal of flexibility for modeling data. If a sufficient numbers of components is allowed, 
the data can be easily modeled with a high degree of accuracy. One of the main abilities of GMM's is 
the capability to form smooth approximations to arbitrarily shaped densities. It can be considered 
as a amalgamation of a uni-modal Gaussian model and a vector quantizer thus enabling us to access 
characteristics of both. Thus the GMM not only provides a smooth overall distribution fit which 
is a feature of uni-modal Gaussian, but also details the multi-modal nature of the density modeled 
by Vector Quantizer.

In our current analysis we sought to use Gaussian Mixture Models to model the probability 
distribution in altitude and thus relate it to the noise generated by the sonar system 
on the UAV. The presence of the sonar on the controller of the UAV induces a change in 
altitude which we try to negate using a direct inversion controller.

The problem is particularly interesting because the behavior of the sonar signal is not directly observable, due to the design of the UAV. Because of this effect we have a deterministic sonar sensor and we need to model it as a stochastic system to determine its output. The noise introduced by the presence of a building from the built-in sonar on the UAV is taken as a random variable.  This is analogous to the automotive industry using a Gaussian random process for characterization of road surfaces \cite{Dodds1973}. The automotive work characterized the shock absorber response of simulated quarter-car models as a random variable \cite{Verros2005}. Like the quarter-car model shock absorbers, the UAV does not have knowledge of upcoming disturbances.  


\section{Experiments and Simulation}


\pgfdeclareimage[width = \textwidth]{testbed}{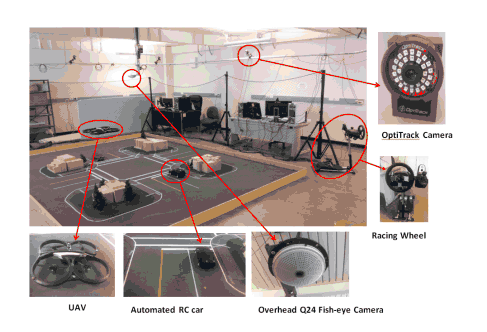}
\pgfdeclareimage[width = 0.46\textwidth]{Xmeanwithpdf}{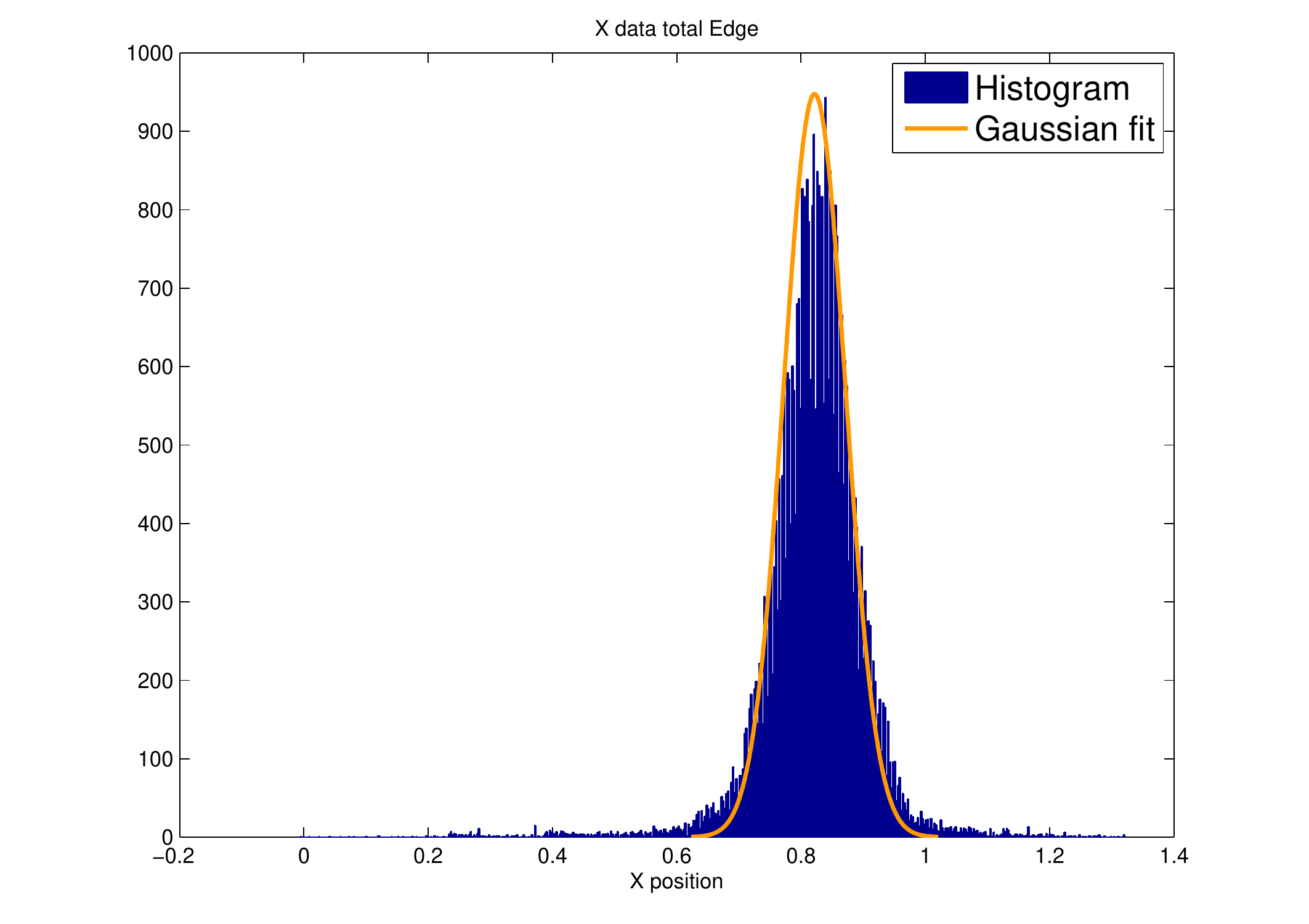}
\pgfdeclareimage[width = 0.46\textwidth]{Ymeanwithpdf}{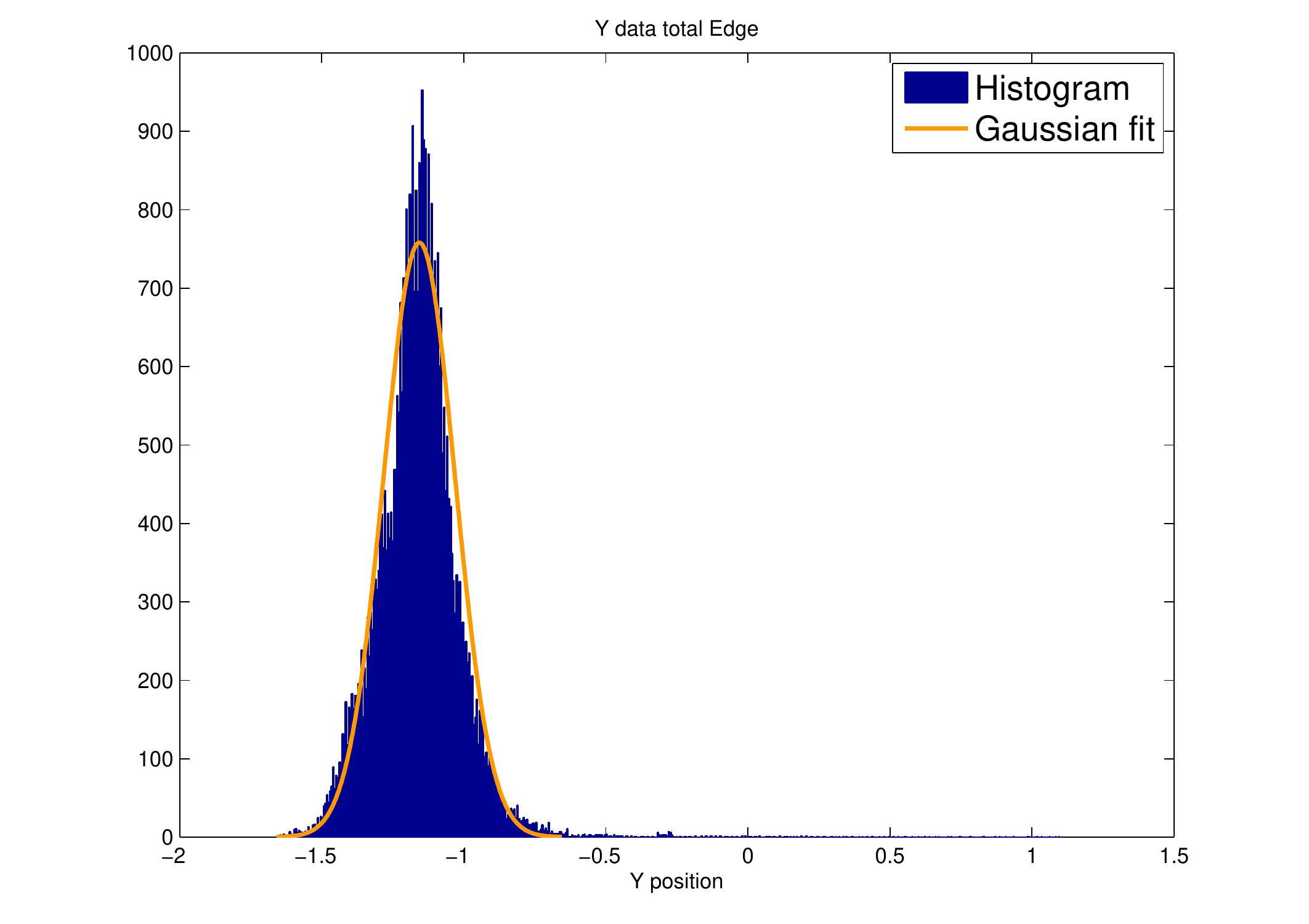}
\pgfdeclareimage[width = 0.46\textwidth]{Zmeanwithpdf}{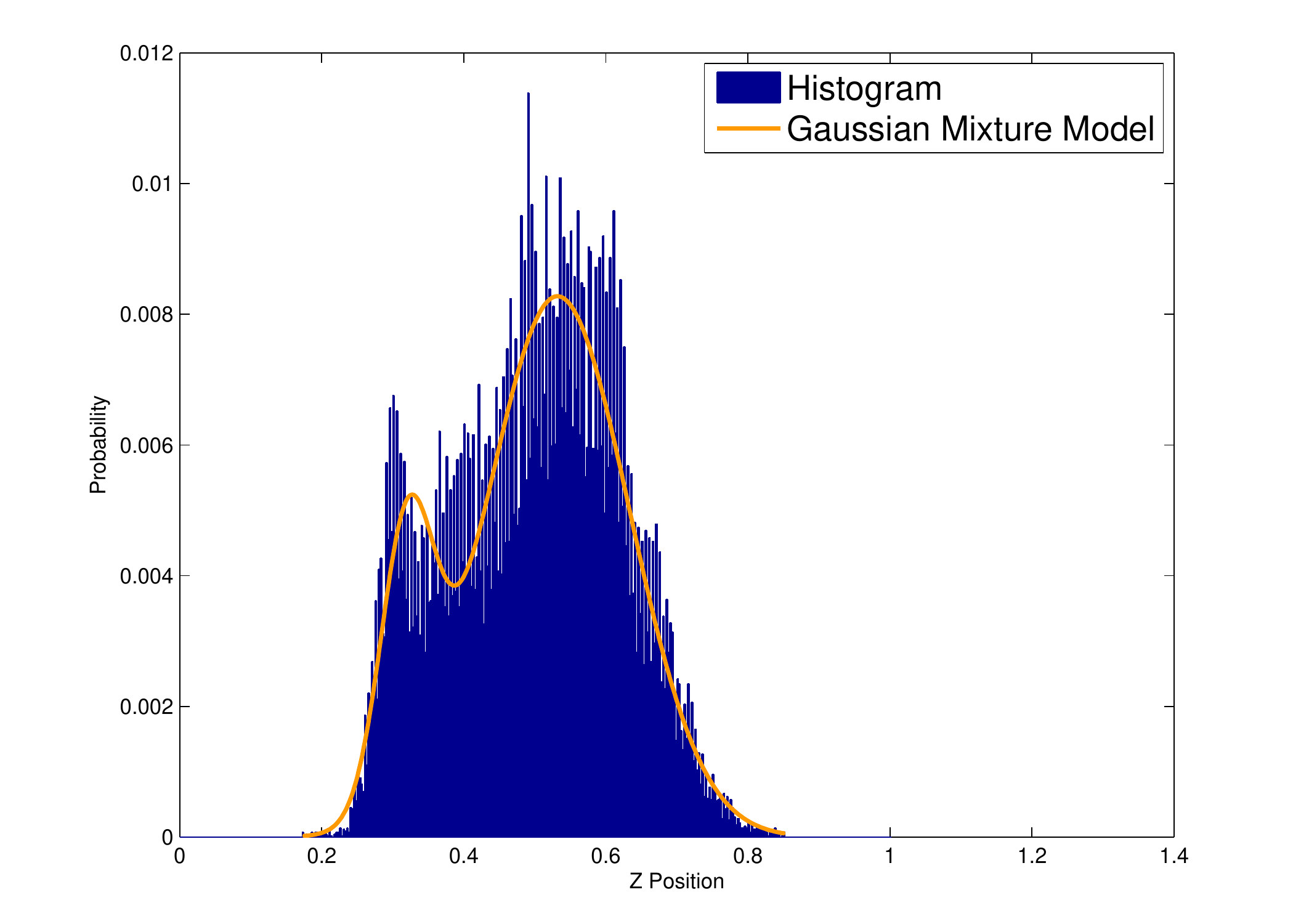}

\begin{figure}
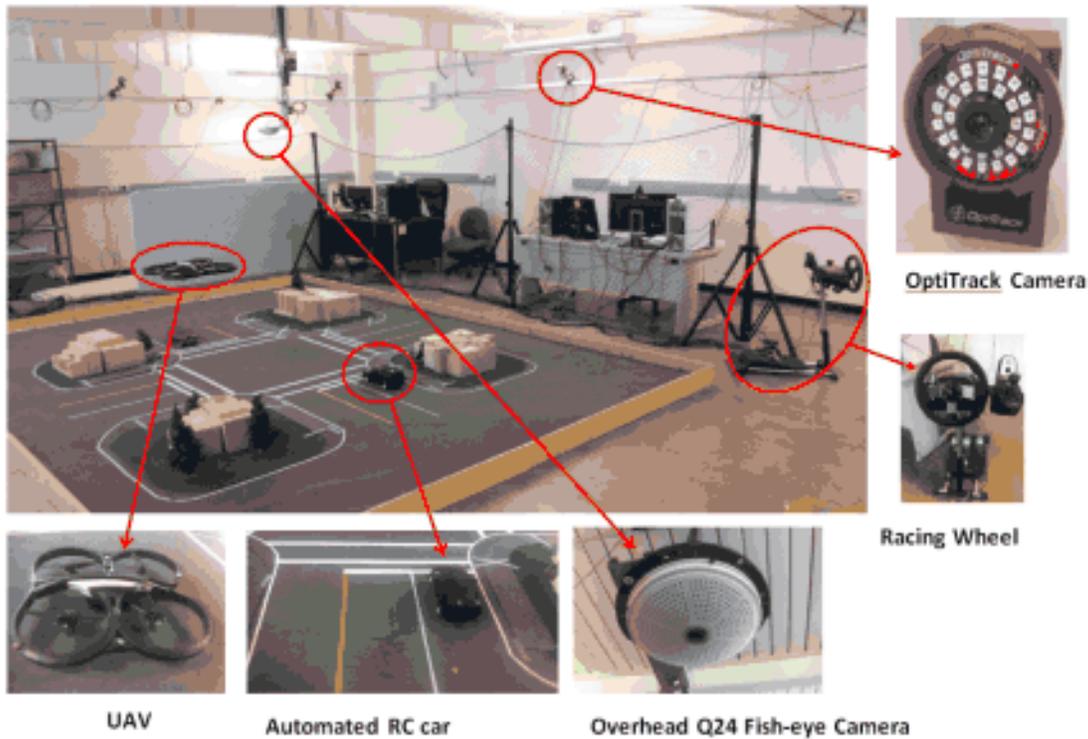

\pgfuseimage{testbed}
\caption{Test Bed}
\label{fig:test bed}
\end{figure}

In our current experimental setup we use an \Otrack{} environment as an external position measurement sensor and place a  UAV in the \Otrack{} environment. The UAV flies in the  experimental testbed shown in Figure \ref{fig:test bed}. The testbed area is $ 16~ft  \times 12~ft$ and it is equipped with the \Otrack{} motion capture system. The \Otrack{} cameras have a optical range between $18~in$ to $433~in$, thus allowing an estimated observable volume of $11 ~ft \times 11 ~ft \times 4 ~ft$.
The \Otrack{} is a set of 10 infra-red cameras, which collect data based on the markers attached to the body.
The UAV used in this experiment is the a AR-Drone Parrot 2.0 Quadrotor (UAV), which is controlled by a client program using the Robot Operating System (ROS)   and our own external controllers. The UAV is designed to receive velocity commands.  The ROS client python script generates the velocity commands and sends the commands to the UAV.  The UAV has two sonar sensors on it's base at a distance of $1 in$, which are programmed to maintain the UAV at a default height of 1 meter.

A single building is placed at a region close to the center. This region is chosen in order to maintain enduring visibility of the UAV from multiple cameras and prevent the \Otrack{} from losing its position. The trajectory of the UAV is designed so that it flies around till it reaches the building and hovers over it until the end of the episodic run. The data is recorded only for those intervals when the UAV is hovering over the building.

A Proportional-Derivative (PD) controller written in Python is used as a closed loop position controller which receives data from the \Otrack{} and calculates the control commands required to maintain the correct position. In the current setup our desired position in altitude is $0.5 m$. 

In our initial experiments we use our baseline PD controller to fly the desired trajectory without the corrective control. Each time the UAV hovers over the building the sonar detects the loss in altitude and sends a signal to the internal on-board controller causing it to rise to a higher altitude. However, this effect pushes the UAV from the building, causing the sonar to mistake the altitude to be too high and thus sending commands to the on-board controller to decrease the altitude, leading to a rapid drop in altitude. 

\begin{figure}
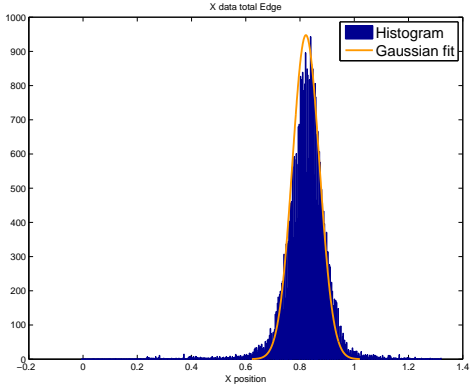
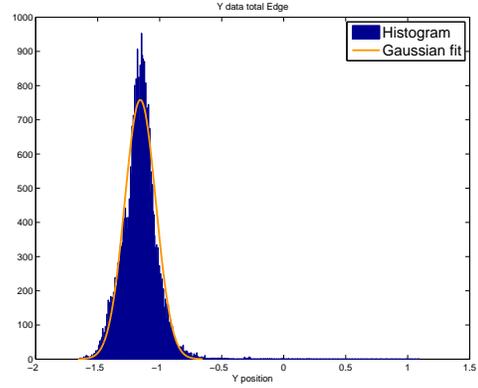
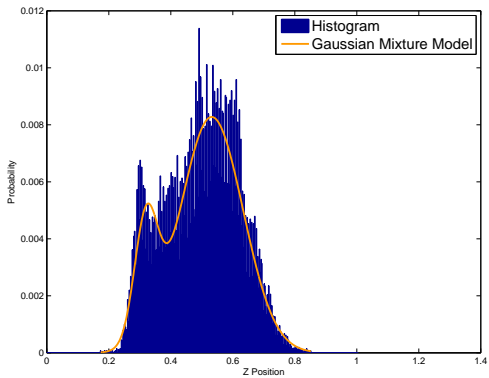

 \subfigure[Hist X]{
    \pgfuseimage{Xmeanwithpdf}
  }
  \subfigure[Hist Y]{
    \pgfuseimage{Ymeanwithpdf}
  }
  \subfigure[Hist Z]{
    \pgfuseimage{Zmeanwithpdf}
  }
  \caption{Histogram distributions}
  \label{Fig:Histogram plots with Gaussian fits}
\end{figure}

We record the position and attitude data and use it to form a distribution of the position along the Z direction. The distribution along all three directions are given in Figure \ref{Fig:Histogram plots with Gaussian fits} a,b,c.

\section{Solution}

Our method is to characterize the noise of the sonar and then cancel it out by updating 
the desired height.  This is an offline method at present, but could be taken to an
online case.  
\subsection{Characterization}
We use recorded data to gain an understanding of the noise.  The UAV
was initially instructed to hover over a flat (no buildings) surface and the x, y, and z position of the craft was 
recorded from the \Otrack{} data.  The UAV was then instructed to hover over a disturbance (the
edge of a building) and the x, y, and z position data was again recorded.  The data
was trimmed to remove obvious outliers as well as takeoff and landing phenomena.

The gathered z-data was placed in bins according to the corresponding
x and y position.  Initially, a polynomial fit of the histogram in Figure \ref{fig:Zmean}. was generated in MATLAB as shown in Figure \ref{fig:Meansurface}.  The intent was to use the polynomial fit to generate a corrective control term to eliminate the effects of the sonar.  The z-data was also used to train a Gaussian process to locally estimate the expected altitude of the UAV and fit the data onto a surface.  

\pgfdeclareimage[width=\textwidth]{Edgemean}{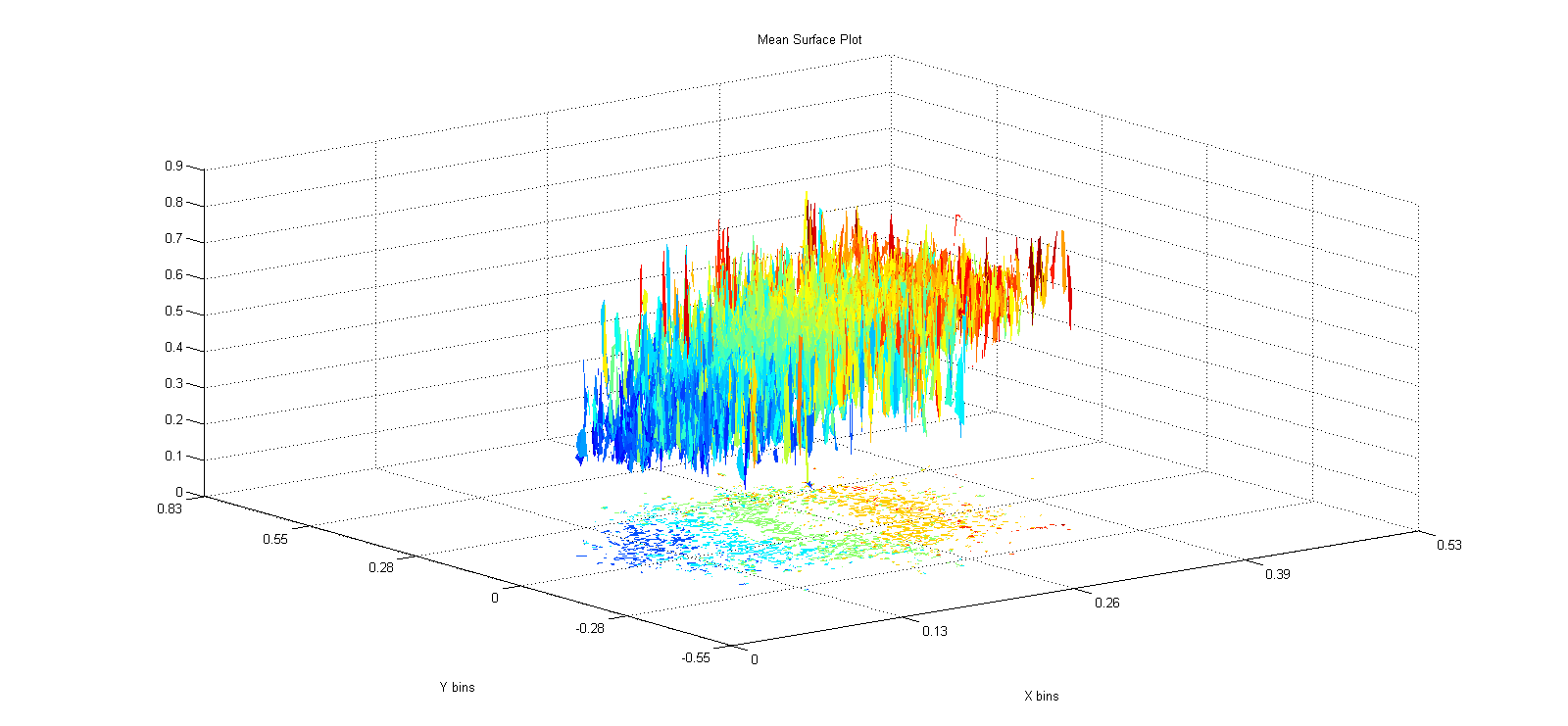}
\begin{figure}
    \centering
    \pgfuseimage{Edgemean}
    \caption{Z Means in each Bin}\label{fig:Zmean}
\end{figure}

\pgfdeclareimage[width = 0.7\textwidth]{Meansurface}{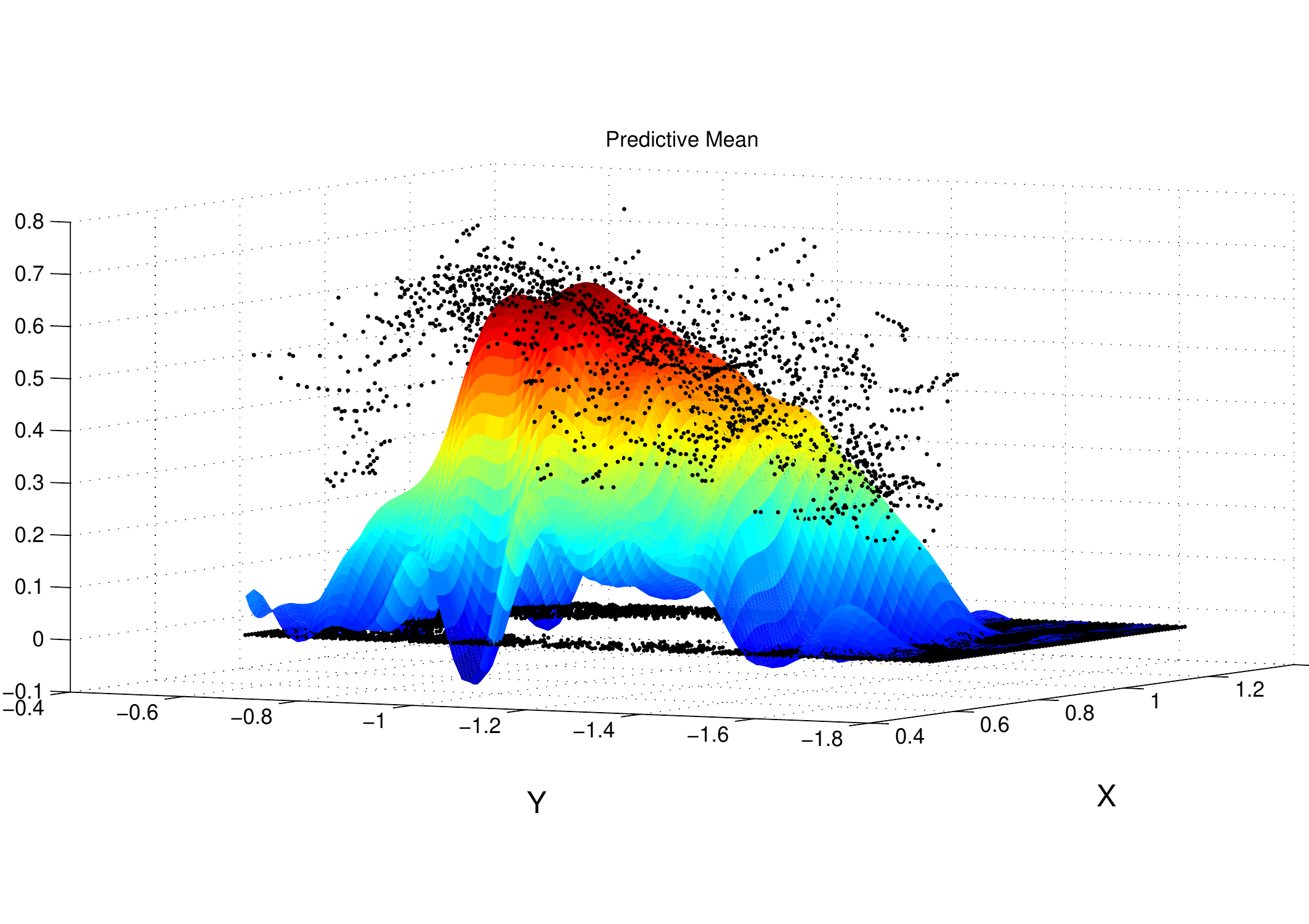}
\begin{figure}
    \centering
    \pgfuseimage{Meansurface}
    \caption{Meansurface}
    \label{fig:Meansurface}
\end{figure}

\subsection{Cancellation} 
The estimates generated by the polynomial surface and the surface generated by the Gaussian Process were not sufficient to eliminate the sonar noise since the sonar instigated both steep and gradual changes in the altitude of the UAV; i.e., a single surface-estimate of the UAV altitude could not anticipate both sonar modes.  A heuristic controller solution was also attempted.  The heuristic controller solution resulted in an insufficient improvement in performance; i.e., the sonar noise was not negated.

The flat terrain data was subtracted from the disturbed data and that result was 
used to form the corrective controller command.  For a given position in the x-y space, the 
desired height command was shifted by the expected value of the noise from the 
sonar in the quadcopter's controller.  This allowed the sonar to add its noise while
not adversely affecting the altitude of the craft.  

\section{Conclusions}
We have used a subset of existing tools at our disposal to model the sonar output in terms of the obtained position data. The random interaction of the sonar with the stepped surface caused conventional statistical modeling techniques to generate intractable estimations of the UAV altitude due to the sonar.  Our results suggest that the problem of eliminating sonar noise is unconventional, and that may warrant the development of a new methodology.  Since the best result was obtained through the heuristic tuning of certain parameters, it is possible that a reinforcement learning approach might be more robust.

\bibliographystyle{ieeetrans}
\bibliography{Stocpro}

\begin{thebibliography}{10}
\providecommand{\url}[1]{#1}
\csname url@samestyle\endcsname
\providecommand{\newblock}{\relax}
\providecommand{\bibinfo}[2]{#2}
\providecommand{\BIBentrySTDinterwordspacing}{\spaceskip=0pt\relax}
\providecommand{\BIBentryALTinterwordstretchfactor}{4}
\providecommand{\BIBentryALTinterwordspacing}{\spaceskip=\fontdimen2\font plus
\BIBentryALTinterwordstretchfactor\fontdimen3\font minus
  \fontdimen4\font\relax}
\providecommand{\BIBforeignlanguage}[2]{{%
\expandafter\ifx\csname l@#1\endcsname\relax
\typeout{** WARNING: IEEEtranS.bst: No hyphenation pattern has been}%
\typeout{** loaded for the language `#1'. Using the pattern for}%
\typeout{** the default language instead.}%
\else
\language=\csname l@#1\endcsname
\fi
#2}}
\providecommand{\BIBdecl}{\relax}
\BIBdecl

\bibitem{Brooks}
R.~Brooks, ``A robust layered control system for a mobile robot.''

\bibitem{Dodds1973}
C.~J. Dodds and J.~D. Robson, ``The description of road surface roughness,''
  \emph{Journal of Sound and Vibration}, vol.~31, no.~2, pp. 175--183, 1973.

\bibitem{Kundur2010}
D.~Kundur, X.~Feng, S.~Liu, T.~Zourntos, and K.~L. Butler-Purry, ``Towards a
  framework for cyber attack impact analysis of the electric smart grid,'' in
  \emph{IEEE Smart Grid Comm}.\hskip 1em plus 0.5em minus 0.4em\relax IEEE,
  October 2010.

\bibitem{Lewis2002}
\BIBentryALTinterwordspacing
J.~A. Lewis, ``Assessing the risks of cyber terrorism, cyber war and other
  cyber threats,'' Center for Strategic and International Studies, Study, 2002.
  [Online]. Available: \url{www.csis.org}
\BIBentrySTDinterwordspacing

\bibitem{Murphy2009}
R.~R. Murphy, ``Mobile robots in mine rescue and recovery,'' \emph{IEEE
  Robotics and Automation Magazine}, vol.~16, pp. 91--103, June 2009.

\bibitem{Murphy2008}
R.~R. Murphy and S.~Stover, ``Rescue robots for mudslides: A descriptive study
  of the 2005 {La Conchita} mudslide response,'' \emph{Journal of Field
  Robotics}, vol.~25, pp. 3--16, Jan 2008.

\bibitem{Permuter}
Permuter, ``A study of gaussian mixture models of color and texture features
  for image classification and segmentation.''

\bibitem{Murphy2010}
S.~S. R.~R.~Murphy, J.~Kravitz, ``Navigational and mission usability in rescue
  robots,'' \emph{Journal of the Robotics Society of Japan}, vol.~28, pp.
  142--146, March 2010.

\bibitem{Reynolds}
D.~Reynolds, ``Gaussian mixture models,'' MIT Lincoln Laboratory, Tech. Rep.,
  2009.

\bibitem{Verros2005}
G.~Verros, S.~Natsiavas, and C.~Papadimitriou, ``Design optimization of
  quarter-car models with passive and semi-active suspensions under random road
  excitation,'' \emph{Journal of Vibration and Control}, vol.~11, pp. 581--606,
  2005.

\end{thebibliography}
\end{document}